\documentclass[12pt]{iopart}
\usepackage{rotating}
\usepackage{dcolumn}
\usepackage{bm}
\usepackage{epsfig}
\usepackage{amssymb}
\usepackage{multirow}
\usepackage{epsfig}

\begin{document}

\title[Understanding jet quenching and medium response with di-hadron correlation]{Understanding jet quenching and medium response with di-hadron correlation}

\author{Jiangyong Jia\dag\
\footnote[3]{jjia@bnl.gov}}
\address{\dag\ Chemistry Department, Stony Brook
University, Stony Brook, NY 11794, USA\\
\dag\ Physics Department, Brookhaven National Laboratory, Upton, NY
11796, USA }

\begin{abstract}
A brief review of the $p_T$ dependence of the dihadron correlations
from RHIC is presented. We attempt to construct a consistent
picture that can describe the data as a whole, focusing on the
following important aspects, 1) the relation between jet
fragmentation of survived jet and medium response to quenched jets,
2) the possible origin of the medium response and its relation to
intermediate $p_T$ physics for single hadron production, 3) the
connection between the near-side ridge and away-side cone, 4) and
their relations to low energy results.
\end{abstract}



%\maketitle
% that are different from $p+p$ or $p+A$ collisions
\section{Introduction}
% dominated either at the near-side by mono-jets emitted from the surface or at the away-side by tangentially emitted back-to-back jets~\cite{Adams:2005ph,Adams:2006yt,Adare:2007vu,Adare:2008cq}.
Dihadron azimuthal correlation has been a successful tool in
understanding the interactions between jet and medium, and in
extracting the properties of the sQGP. Over the years, the
correlation analyses have been carried out in various regions of
transverse momentum ($p_T$) for the triggers and partners. Many
interesting features have been discovered. In the high $p_T$
region, the correlation distributions show narrow peaks around
$\Delta\phi\sim0$ (near-side) and $\Delta\phi\sim\pi$
(away-side)~\cite{Adams:2005ph,Adams:2006yt,Adare:2007vu,Adare:2008cq},
consistent with fragmentation of jets escaping the dense medium
with small energy loss. In the low $p_T$ region, the correlation
distributions are dominated by a double hump structure around
$\Delta\phi\sim\pi\pm1.1$ at the away-side (the
cone)~\cite{Adams:2005ph,Adler:2005ee,Horner:2007gt,
Adare:2007vu,Adare:2008cq} and a structure elongated along the
$\Delta\eta$ but centered around $\Delta\phi\sim0$ (the
ridge)~\cite{Adams:2006tj,Adare:2008cq}, characteristic of a
complicated response of the medium to energy deposited by the
quenched jets.

In the meanwhile, many theoretical models~\cite{theory} have been
proposed to interpret the data. But to date, a complete and
consistent picture accommodating the vast amount data is still
missing. Our goal is to provide a brief overview of the dihadron
correlation results, with an eye towards the reciprocal relation
between jet quenching and medium response, and discuss several
insights distilled from the data.

\section{Correlation landscape in $p_T^A$ and $p_T^B$}

\begin{figure}
\begin{center}
\epsfig{file=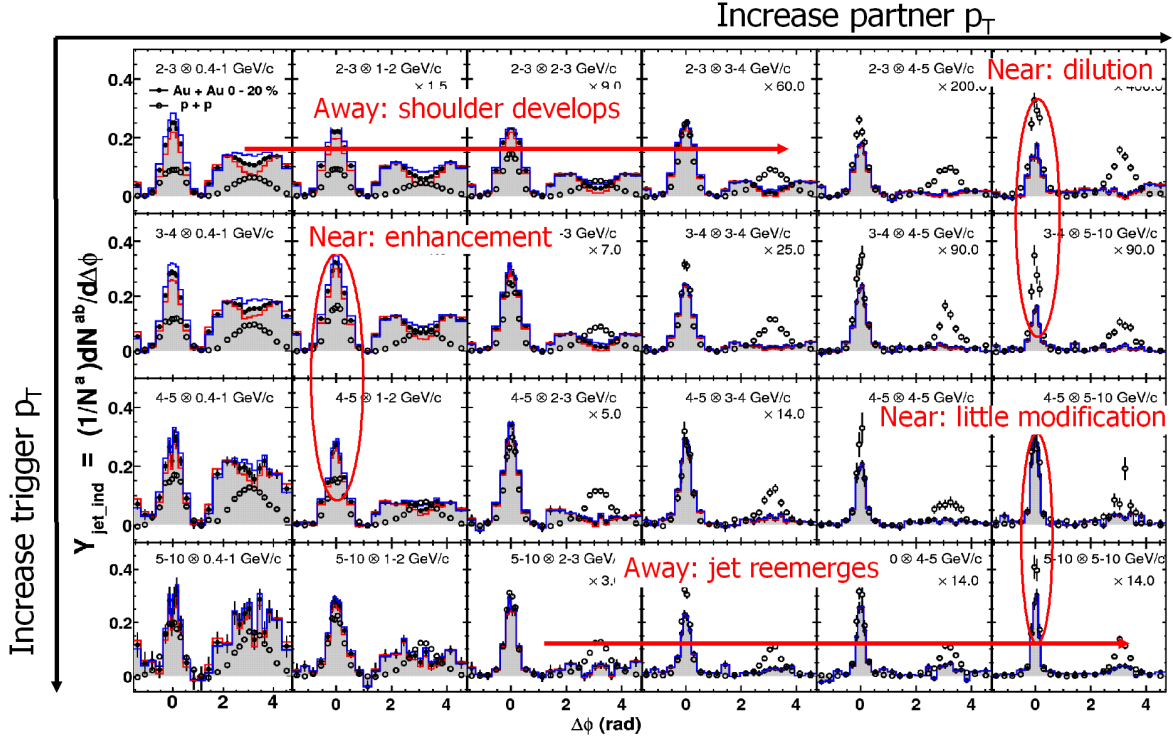,width=1\columnwidth}
\caption{\label{fig:1a} The $\Delta\phi$ distribution in fine bin of trigger and partner $p_T$~\cite{Adare:2008cq}. Several important features are indicated by the lines and circles.}
\end{center}
\end{figure}

\begin{figure}[h]
\begin{tabular}{lr}
\begin{minipage}{0.45\linewidth}
\begin{flushleft}
\epsfig{file=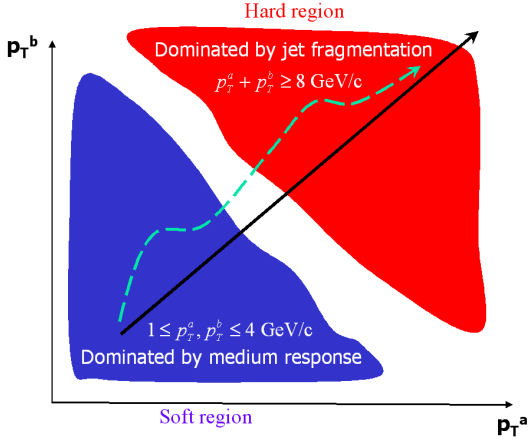,width=1\columnwidth}
\end{flushleft}
\end{minipage}
&
\hspace*{-0.5in}\begin{minipage}{0.60\linewidth}
\begin{flushright}
\caption{\label{fig:1b} A schematic view of the $p_T$ dependence of the dihadron correlation (applicable for both near- and away-side). The $p_T^A\otimes p_T^B$ region dominated by jet fragmentation (top right region)
and by medium response (bottom left) are indicated. Arrows indicate the possible routes for scanning from low to high $p_T$.}
\end{flushright}
\end{minipage}
\end{tabular}
\end{figure}

In general the dihadron correlations depend on the $p_T$ of both
hadrons in the pair, and the full characterization of their
modification patterns have to be studied differentially as function
of trigger $p_T$ ($p_T^A$) and partner $p_T$ ($p_T^B$). Such a
survey study has been carried out recently by the
PHENIX~\cite{Adare:2008cq} and STAR
Collaboration~\cite{Horner:2007gt}. Fig.~\ref{fig:1a} summarizes
dihadron $\Delta\phi$ distribution in a broad transverse momentum
space, which shows many distinctive features appearing at different
$p_T$ regions (indicated by the circles and lines). These features
fits well into a simple two-component picture as illustrated in
Fig.~\ref{fig:1b} separately for both the near- and away-side: a
jet fragmentation component that dominates for
$p_T^A+p_T^B\gtrsim8$ GeV/$c$, and a medium response component that
dominates at $p_T^A,p_T^B<4$ GeV/$c$. The rich $p_T$ dependent
correlation patterns simply reflect the competition between
fragmentation of survived jets and medium response to quenched jets
on both the near- and away-side. The observed patterns are rather
complicated in Fig.~\ref{fig:1a} since 1) the medium response and
jet fragmentation have very different angular distribution and very
different spectral slope, 2) the shapes of the medium response are
also quite different between the near- and away-side.

%The question is not the shape, but the enhancement of the shoulder region.
%PHENIX data provide a true comparison to p+p data, instead of the dAu which might suffer from the initial state effects.

\section{Medium response}
%~\footnote{There is an approximate scaling on the away-side, since the jet energy also depends on away-side hadron $p_T$(seesaw effects).}

\begin{figure}[h]
\begin{center}
\epsfig{file=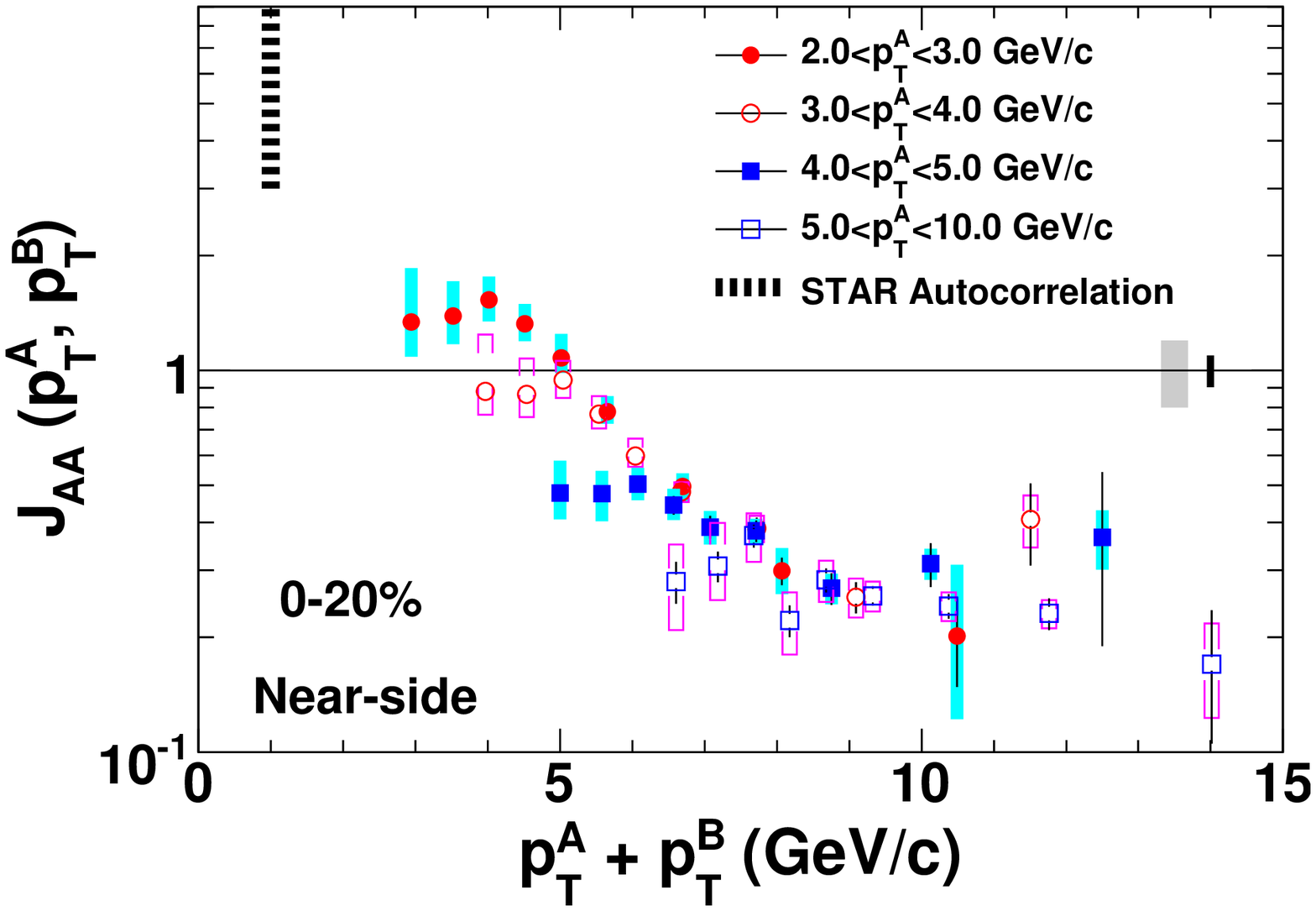,width=0.4\columnwidth}
\epsfig{file=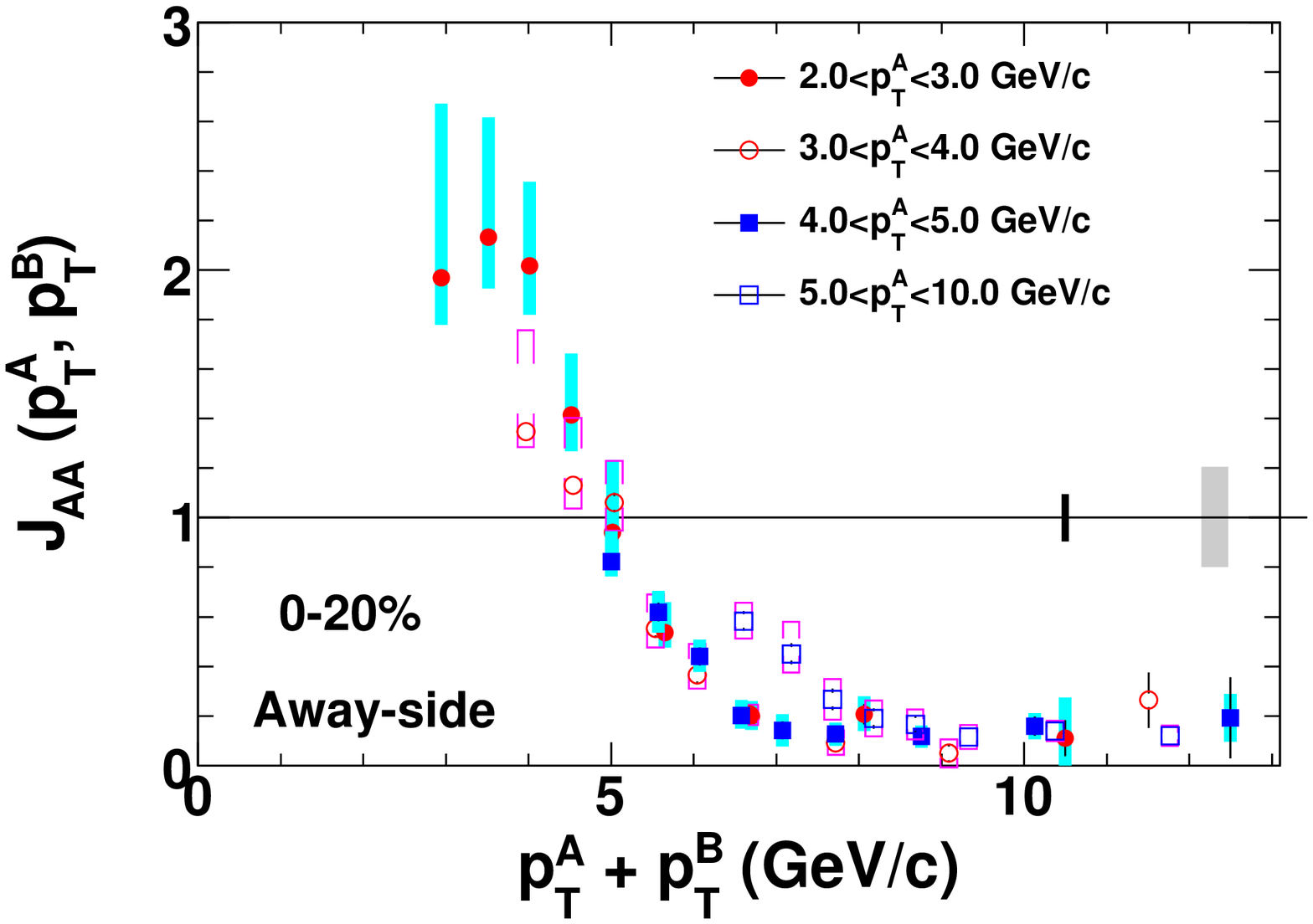,width=0.4\columnwidth}
\caption{\label{fig:2} The modification factor for hadron pair yield as function of $p_T^{sum} = p_T^A+p_T^B$ for the near-side (left) and away-side (right).
$p_T^{sum}$ condenses the 2-D correlation data in $p_T^A$ and $p_T^B$
space into a one dimensional plot. The STAR auto-correlation result~\cite{md} is divided by 3 (the lower end) to normalize the $\eta$ acceptance relative to PHENIX.}
\end{center}
\end{figure}

A new variable $J_{AA}$ was introduced recently to describe the
medium response at low $p_T$~\cite{Adare:2008cq}. $J_{AA}$ quantify
the medium modification of hadron pair yield from the expected
yield, in a way similar to $R_{AA}$ for describing the modification
of single hadron yield. The hadron pair yield is proportional to
the dijet yield, and in the absence of nuclear effects, it should
scale with $N_{\rm coll}$, and $J_{AA}=1$. Fig.~\ref{fig:2} shows
$J_{AA}$ as a function of pair proxy energy ($p_T^{sum} =
p_T^A+p_T^B$) for the near- (left panel) and away-side (right
panel). The STAR autocorrelation result~\cite{md} is shown as a
single point at $2\langle p_T\rangle\sim1$ GeV/$c$. In contrast to
a constant suppression at large $p_T^{sum}$, the pair yields are
not suppressed or even enhanced at $p_T^{sum}<8$ GeV/$c$. This
enhancement directly reflects the energy transport that
redistribute energy of the quenched jets to low $p_T$ hadrons
(medium response). We would like to point out that $p_T^{sum}$ is a
natural variable for the near-side correlation since it
approximates the jet energy, and data show an approximate scaling
in $p_T^{sum}$. However, even the data points for the away-side
tend to group together, probably because the medium response is a
function of jet energy, which increase monotonously with
$p_T^{A,B}$.

\begin{figure}
\begin{center}
\epsfig{file=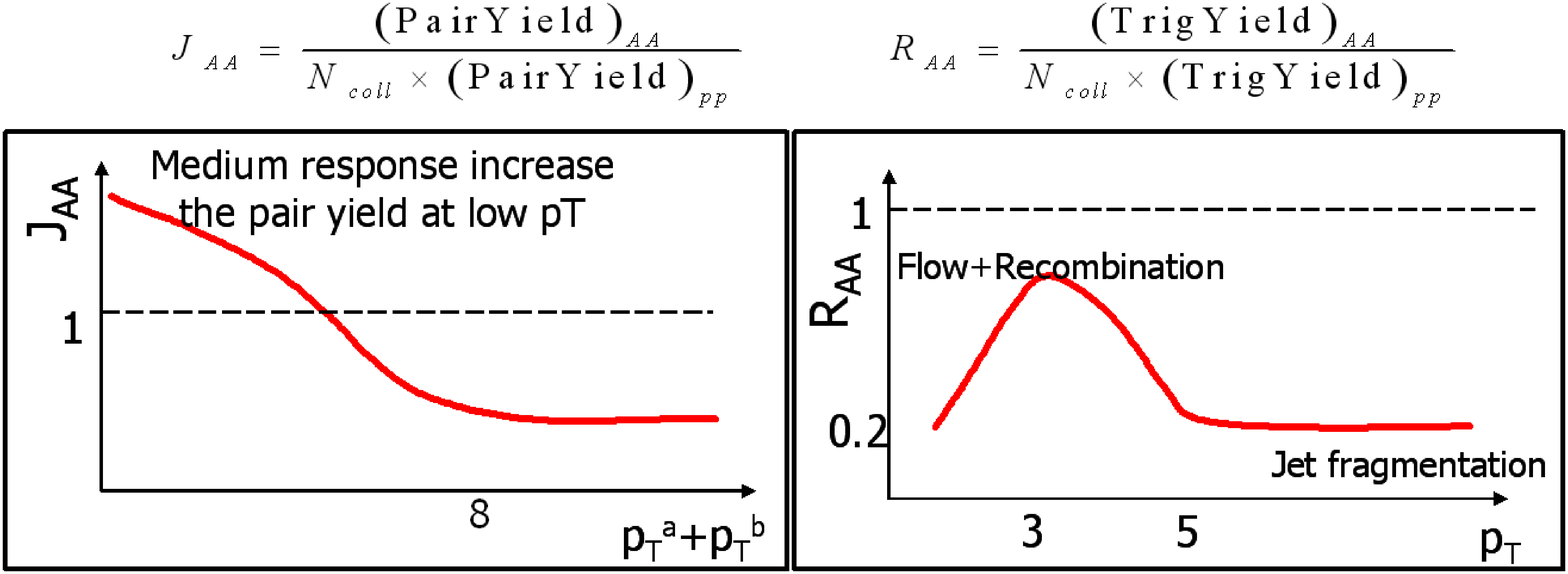,width=0.7\columnwidth}
\caption{\label{fig:3} Schematic view of the $p_T$ dependence modification for single hadrons (right panel) and hadron pairs (left panel).}
\end{center}
\end{figure}

The transition from jet fragmentation dominated to medium dominated
region in dihadron correlation happens around $p_T\sim4-5$ GeV/$c$,
a region similar to that for the single particle from soft physics
(hydrodynamics+ recombination) dominated region to hard physics
(jet) dominated region. Naturally, we expect the physics important
for single particle production should play an important role for
the dihadron correlation. Fig.~\ref{fig:3} shows schematically the
$p_T$ dependence of the modifications of the single particle yield
(via $R_{AA}$) and hadron pair yield (via $J_{AA}$). Their $p_T$
dependence trend are quite different, especially at low $p_T$,
which can be explained qualitatively as follows. Even though jet
production dominates single particle yield at $p_T>2$ GeV/$c$ in
$p+p$ collisions, the strong energy loss and collective flow modify
the $p_T$ distribution by shifting hard hadrons to lower $p_T$ and
pushing soft hadrons to higher $p_T$. This reshuffling changes
single-hadron and correlated hadron-pair yield, hence the $R_{AA}$
and $J_{AA}$ shape. Indeed, several theoretical models suggest that
collective flow and recombination play a significant role in
modifying the angular shape, spectra slope and particle composition
of the correlated pairs~\cite{theory}. $J_{AA}$ provides a mean to
quantify the contribution of jet fragmentation hadrons or jet
induced hadrons in this $p_T$ region.

\section{Dilution of per-trigger yield by non-fragmentation hadrons}

\begin{figure}
\begin{center}
\epsfig{file=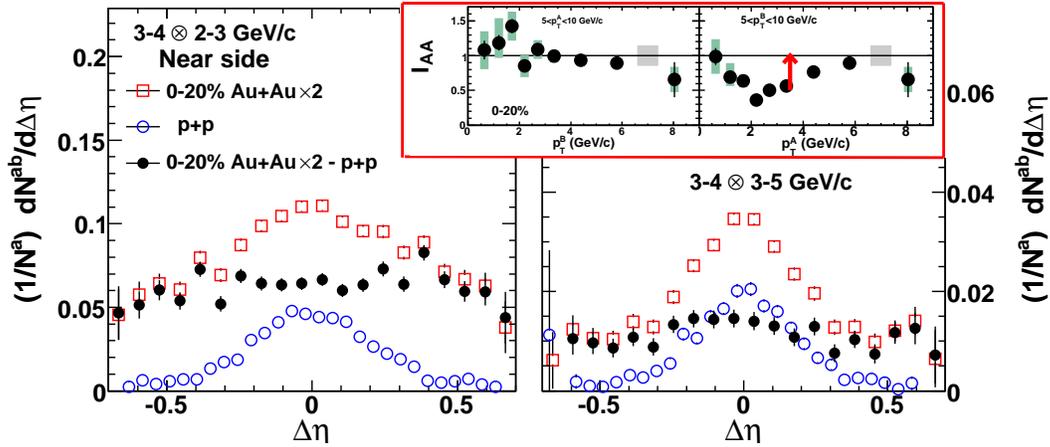,width=0.9\columnwidth}
\caption{\label{fig:4} Per-trigger yield $\Delta\eta$ distribution for 3-4 GeV/$c$ triggers and two parter $p_T$ selections. The ridge distribution (solid circles) is estimated by subtracting the Au+Au distribution corrected by dilution effect (open squares) minus the p+p (open circles).
The dilution correction ($\times2$) is indicated by the red arrow in the inserted panel (see text for explanation).}
\end{center}
\end{figure}

Previously, the modification of dihadron yield is characterized
with $I_{AA}$ (ratio of per-trigger yield between Au+Au and
$p+p$)~\cite{Adler:2002tq}. $I_{AA}$ is a good variable at high
$p_T$, since most triggers come from jets and most jets fragment
into at most one trigger, such that per-trigger yield (PTY) is a
good representation of per-jet yield. However at lower $p_T$
region, non-fragmentation triggers from soft production mechanisms
or medium response mechanisms become important. These triggers tend
to dilutes the $I_{AA}$, since they either has no correlation or
non-jet like correlation (such as ridge). Fig.~\ref{fig:4}
illustrate the dilution effects for near-side $\Delta\eta$
correlation. We estimate dilution factor ($\sim2$) for 3-4 GeV/$c$
triggers based on their correlations with 5-10 GeV/$c$ hadrons as
shown by the inserted panel: requiring 5-10 GeV/$c$ hadrons ensures
the pairs are dominated by the jet fragmentation (left panel of the
insert), thus deviation of $I_{AA}$ from one for soft triggers
reflects the level of dilution (the red arrow). Once the dilution
factor is corrected, we subtract out the jet fragmentation
contribution and obtain the ridge distribution (black circles). The
estimated ridge contribution is approximately flat, consistent
experimental data at large $\Delta\eta$. However, this dilution
effect was not observed in some STAR
analysis~\cite{Bielcikova:2007mb, Putschke:2007mi}, which shows
that the PTY$_{AA}$ subtracted by the estimated ridge before any
dilution correction already equals PTY$_{pp}$.

\section{Origin of correlated pair and connection between near-side ridge and away-side cone}
In most correlation analyses and model calculations, it was
normally assumed that one hadron (``trigger'') comes from the jet,
and the second hadron (``partner'') comes from either fragmentation
or feedback, i.e. only jet-jet and jet-medium pairs are considered.
In this picture, the trigger comes from a jet that is biased to the
surface, which losses some energy and fragments outside the medium.
The fragments contribute to the near-side jet peak, and the
feedback of the lost energy gives rise to the near-side ridge. In
parallel, the away-side jet is quenched as it traverses a longer
medium, contributing to the away-side cone.

\begin{figure}
\begin{center}
\epsfig{file=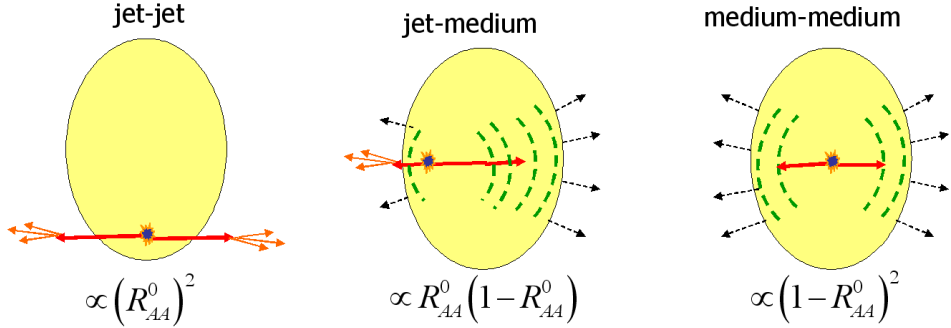,width=0.8\columnwidth}
\caption{\label{fig:5} The geometrical origin of the jet-jet (left), jet-medium (middle) and medium-medium (right) pairs. Their production rate scale with the volume of the emission points,
which is proportional to $(R^0_{AA})^2$,$R^0_{AA}(1-R^0_{AA})$, and $(1-R^0_{AA})^2$, respectively ($R^0_{AA}$ is the constant suppression level at high $p_T$.)}
\end{center}
\end{figure}

This picture does not include the medium-medium pairs (both hadrons
come from medium feedback of quenched jets). These pairs could be
important at intermediate and low $p_T$, since each jet can induce
more correlated hadrons via jet quenching than via fragmentation.
For example, most medium response models induce correlation by
local heating of medium by the jet, such as momentum kick, jet
deflection, mach-cone, etc~\cite{theory}, which are very effective
in generating large yield of correlated hadron pairs. In addition,
the whole overlap volume contributes to the observed medium-medium
pairs, while both jet-jet and jet-medium pairs suffer a strong
suppression. This point is illustrated by Fig.~\ref{fig:5}, which
shows the typical geometrical origin for the three types of
correlated pairs. The jet fragmentation contribution is
proportional to the number of survived jet ($\propto R^0_{AA}$,
i.e. the constant suppression level at large $p_T$,
$R^0_{AA}\approx0.2$ in most central bin.), while the medium
response is proportional to the number of quenched jets ($\propto
1-R^0_{AA}$). For jet-jet pairs, both hadron are emitted from the
surface (tangential emission); for jet-medium pairs, the jet hadron
is emitted from the surface (surface emission) and the other from
the whole volume; for medium-medium pairs, both hadrons are emitted
from the whole volume. The production rate for jet-jet, jet-medium
and medium-medium pairs scale approximately with $(R^0_{AA})^2$,
$R^0_{AA}(1-R^0_{AA})$ and $(1-R^0_{AA})^2$. Clearly, if
$R^{0}_{AA}\rightarrow 0$, the medium-medium pairs becomes
dominant.

Recently, several analyses have been carried out to quantify the
properties of the near-side ridge and away-side cone
structures~\cite{Bielcikova:2007mb,Putschke:2007mi,Putschke:2007cb,Afanasiev:2007wi,af,Adare:2008cq}.
The data show very similar properties between the ridge and the
cone, i.e. both have similar slope and bulk like particle
compositions, and both are important up to 4 GeV/$c$. These
similarities suggests that their production mechanisms are
connected. The medium-medium pairs from quenched jets are natural
candidates for creating these similarities. Because medium-medium
pairs come from quenched jets originated deep inside the medium,
they contribute to both the near-side and away-side on a equal
footing. The near-side pairs could contain correlations among mach
cone particles, and away-side pairs could also contain correlation
between the ridge and mach cone particles (see
Ref.~\cite{Jia:2008vk} for a possible realization).

\section{Energy dependence and three-particle correlation}

\begin{figure}[h]
\begin{tabular}{cl}
\begin{minipage}{0.7\linewidth}
\begin{flushleft}
\epsfig{file=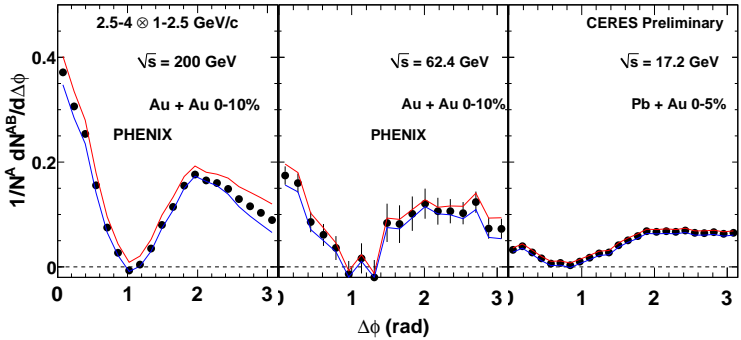,width=0.9\columnwidth}
\end{flushleft}
\end{minipage}
&
\hspace*{-0.9in}\begin{minipage}{0.4\linewidth}
\begin{flushleft}
\caption{\label{fig:6} The $\Delta\phi$ distribution in central collisions for three collision energies in central collisions.}
\end{flushleft}
\end{minipage}
\end{tabular}
\end{figure}

\begin{figure}
\begin{center}
\epsfig{file=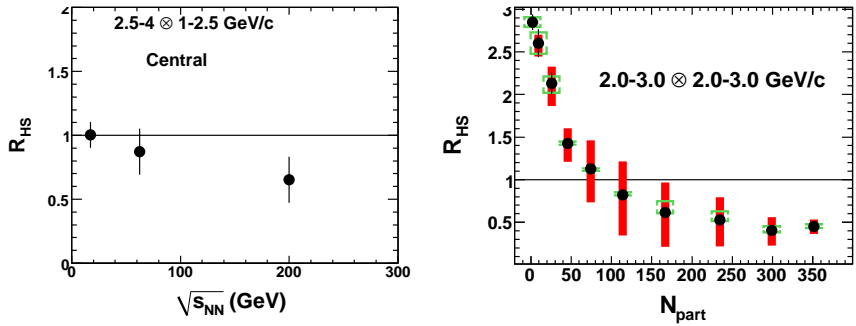,width=0.75\columnwidth}
\caption{\label{fig:7} (Left) The $\sqrt{s}$ dependence of the $R_{HS}$ calculated from Fig.\ref{fig:6}. (Right) centrality dependence of $R_{HS}$ for a lower $p_T$ bin
in $\sqrt{s_{NN}}=200$ GeV Au+Au collisions~\cite{Adare:2008cq}.}
\end{center}
\end{figure}

A strong modification of the away-side correlation was also
observed at the top SPS energy ($\sqrt{s_{NN}}=17.2$
GeV)~\cite{Ploskon:2007es}. The strong away-side broadening has
been used to argue for a similar interpretation (such as mach cone)
as for the RHIC results. However a quantitative analysis of the
energy dependence of the modification patterns (see
Fig.\ref{fig:6}) shows that the yield of medium response are quite
different between RHIC and SPS energies. In fact the near-side
yield drop by almost factor of 8 going from 200 GeV to 17 GeV while
the away-side shoulder yield drops by factor of 2 in the same
energy range. But there are little dependence of the yields on
$\sqrt{s}$ in the away-side head region, where the jet
fragmentation is important. To quantify the energy dependence of
away-side shape, we calculate the ratio of the yield density in the
head region to that in shoulder region,
$R_{HS}$~\cite{Adare:2007vu}, in Fig.\ref{fig:7}a. The $R_{HS}$
increases with decreasing collision energy, with a ratio slightly
above one in SPS energy. This value is comparable with that
obtained for rather peripheral ($N_{part} = 70$) in Au+Au
collisions at 200 GeV (Fig.\ref{fig:7}b). These results suggest a
much weaker medium response in SPS energy (the ridge almost
disappeared and cone strongly suppressed) than that at RHIC, but a
similar jet fragmentation contribution, probably related to smaller
energy loss and stronger Cronin effects at lower
energy~\cite{adare:2008cx}.

Lastly, it was shown that the mach-cone angle found from the three
particle (3-p) correlation (1.4 rad) is significantly larger than
the two particle (2-p) correlation analysis (1.1 rad). SPS also
seems to see a 3-p correlation signal~\cite{3p}. However one should
realized that, it is possible that the kinematics of jets
contributing to 3-p signal is different from those contributing to
the 2-p signal. It is likely that most jet have multiplicity $<3$
that the 3-p only samples a small fraction of all jets that
contribute to the dihadron correlations.

\section{Discussion}
Due to the surface bias and steeply falling parton spectra, the
observed high $p_T$ single hadrons and dihadron pairs mainly come
from those jets that suffer minimal interaction with the medium.
This energy loss bias limits their usefulness as tomography tools.
On the other hand, medium responses are directly sensitive to the
energy loss and energy dissipation processes used to model the high
$p_T$ production. For example the collisional energy loss would
imply that momentum kick dominates the low $p_T$ pairs, the
radiative energy loss would favor for the gluon feedback mechanism.
Finally, the jet quenching and medium responses are modeled
separately in most theoretical calculations. A unified framework,
including both jet quenching and medium response, which can
describe the full $p_{T}$ evolution of the jet shape and yield at
both near- and away-side is required to understand the details of
the parton-medium interactions.


\section*{References}
\begin{thebibliography}{99}
\bibitem{Adams:2005ph}
  J.~Adams {\it et al.} [STAR Collaboration],
  %``Distributions of charged hadrons associated with high transverse  momentum
  %particles in p p and Au + Au collisions at s(NN)**(1/2) =  200-GeV,''
  Phys.\ Rev.\ Lett.\  {\bf 95}, 152301 (2005)
  %arXiv:nucl-ex/0501016.
%\cite{Adams:2006yt}
\bibitem{Adams:2006yt}
  J.~Adams {\it et al.}  [STAR Collaboration],
  % ``Direct observation of dijets in central Au + Au collisions at s(NN)**(1/2)
  %= 200-GeV,''
  Phys.\ Rev.\ Lett.\  {\bf 97}, 162301 (2006)
\bibitem{Adare:2007vu}
  A.~Adare {\it et al.}  [PHENIX Collaboration],
  %``Transverse momentum and centrality dependence of dihadron correlations   in
  %Au+Au collisions at sqrt(s_NN)=200 GeV: Jet-quenching and the response of
  %partonic matter,''
  Phys.\ Rev.\ C {\bf 77}, 011901(R) (2008)
\bibitem{Adare:2008cq}
  A.~Adare {\it et al.}  [PHENIX Collaboration],
  %``Dihadron azimuthal correlations in Au+Au collisions at sqrt(s_NN)=200
  %GeV,''
  arXiv:0801.4545 [nucl-ex].

%\cite{Horner:2007gt}
\bibitem{Horner:2007gt}
  M.~J.~Horner  [STAR Collaboration],
  %``Low- and intermediate-p(T) di-hadron distributions in Au + Au collisions at
  %s(NN)**(1/2) = 200-GeV from STAR,''
  J.\ Phys.\ G {\bf 34}, S995 (2007)
%  [arXiv:nucl-ex/0701069].
\bibitem{Adler:2005ee}
  S.~S.~Adler {\it et al.} [PHENIX Collaboration],
  %``Modifications to di-jet hadron pair correlations in Au + Au collisions at s(NN)**(1/2) = 200-GeV,''
  Phys.\ Rev.\ Lett.\  {\bf 97}, 052301 (2006)
  %arXiv:nucl-ex/0507004.
%\cite{Adams:2006tj}
\bibitem{Adams:2006tj}
  J.~Adams {\it et al.}  [STAR Collaboration],
  %``Delta(phi) Delta(eta) correlations in central Au + Au collisions at
  %s(NN)**(1/2) = 200-GeV,''
  Phys.\ Rev.\  C {\bf 75}, 034901 (2007)
  %  [arXiv:nucl-ex/0607003].

\bibitem{theory}
  N.~Armesto, C.~A.~Salgado and U.~A.~Wiedemann,
  Phys.\ Rev.\ Lett.\  {\bf 93}, 242301 (2004),
  Phys.\ Rev.\  C {\bf 72}, 064910 (2005);
  %arXiv:hep-ph/0411341.
  I.~Vitev,
  %``Large angle hadron correlations from medium-induced gluon radiation,''
  Phys.\ Lett.\ B {\bf 630}, 78 (2005);
  I.~M.~Dremin,
  JETP Lett.\  {\bf 30}, 140 (1979)
  %[Pisma Zh.\ Eksp.\ Teor.\ Fiz.\  {\bf 30}, 152 (1979)].
  V.~Koch, A.~Majumder and X.~N.~Wang,
  Phys.\ Rev.\ Lett.\  {\bf 96}, 172302 (2006)
  J.~Casalderrey-Solana, E.~V.~Shuryak and D.~Teaney,
  hep-ph/0602183.
  T.~Renk and J.~Ruppert,
  %``Mach cones in an evolving medium,''
  Phys.\ Rev.\  C {\bf 73}, 011901 (2006);
  C.~B.~Chiu and R.~C.~Hwa,
  %``Away-side azimuthal distribution in a Markovian parton scattering model,''
  Phys.\ Rev.\ C {\bf 74}, 064909 (2006);
  %[arXiv:nucl-th/0609038].
  A.~D.~Polosa and C.~A.~Salgado,
  Phys.\ Rev.\  C {\bf 75}, 041901(R)(2007);
  P.~Romatschke,
  %``Momentum broadening in an anisotropic plasma,''
  Phys.\ Rev.\  C {\bf 75}, 014901 (2007);
  %[arXiv:hep-ph/0607327].
  A.~Majumder, B.~Muller and S.~A.~Bass,
  %``Longitudinal Broadening of Quenched Jets in Turbulent Color Fields,''
  Phys.\ Rev.\ Lett.\  {\bf 99}, 042301 (2007);
  C.~Y.~Wong,
  %``Ridge Structure associated with the Near-Side Jet in the (Delta phi)-(Delta
  %eta) Correlation,''
  Phys.\ Rev.\  C {\bf 76}, 054908 (2007);
  E.~V.~Shuryak,
  %``On the Origin of the 'Ridge' phenomenon induced by Jets in Heavy Ion
  %Collisions,''
  Phys.\ Rev.\  C {\bf 76}, 047901 (2007);
  %[arXiv:0706.3531 [nucl-th]].
  V.~S.~Pantuev,
  %``'Jet-Ridge' effect in heavy ion collisions as a back splash from stopped
  %parton,''
  arXiv:0710.1882 [hep-ph];

\bibitem{md} M.~Daugherity [STAR Collaboration], QM2008 talk, see
    slide 16, and these
    proceedings.
\bibitem{Adler:2002tq}
  C.~Adler {\it et al.}  [STAR Collaboration],
  %``Disappearance of back-to-back high p(T) hadron correlations in central Au +
  %Au collisions at s(NN)**(1/2) = 200-GeV,''
  Phys.\ Rev.\ Lett.\  {\bf 90}, 082302 (2003)

%\cite{Bielcikova:2007mb}
\bibitem{Bielcikova:2007mb}
  J.~Bielcikova,
  %``Azimuthal and pseudo-rapidity correlations with strange particles at
  %intermediate-p(T) at RHIC,''
  J.\ Phys.\ G {\bf 34}, S929 (2007)
  %[arXiv:nucl-ex/0701047].
%\cite{Putschke:2007mi}
\bibitem{Putschke:2007mi}
  J.~Putschke,
  %``Intra-jet correlations of high-$p_t$ hadrons from STAR,''
  J.\ Phys.\ G {\bf 34}, S679 (2007), see Fig.5.

%\cite{Putschke:2007cb}
\bibitem{Putschke:2007cb}
  J.~Putschke,
  %``Near-side Delta(eta) correlations of high-p(t) hadrons from STAR,''
  Eur.\ Phys.\ J.\  C {\bf 49}, 57 (2007).
%\cite{Afanasiev:2007wi}
\bibitem{Afanasiev:2007wi}
  S.~Afanasiev {\it et al.}  [PHENIX Collaboration],
  %``Particle-species dependent modification of jet-induced correlations in
  %Au+Au collisions at sqrt(s_NN) = 200 GeV,''
  arXiv:0712.3033 [nucl-ex].
\bibitem{af} A.~Frantz, [PHENIX Collaboration] these proceedings.
\bibitem{Jia:2008vk}
  J.~Jia and R.~Lacey,
  %``Influence of quenched jets on di-hadron correlations,''
  arXiv:0806.1225 [nucl-th].
\bibitem{Ploskon:2007es}
  M.~Ploskon  [CERES Collaboration],
  %``Two particle azimuthal correlations at high transverse momentum in Pb - Au
  %at 158-AGeV/c,''
  Nucl.\ Phys.\  A {\bf 783}, 527 (2007)
  %[arXiv:nucl-ex/0701023].
\bibitem{adare:2008cx}
  A.~Adare {\it et al.}  [PHENIX Collaboration],
  %``Energy dependence of pi-zero production in Cu+Cu collisions at sqrt(s_NN) =
  %22.4, 62.4, and 200 GeV,''
  arXiv:0801.4555 [nucl-ex].
\bibitem{3p} J.~G.~Ulery, these proceedings.
\end{thebibliography}
\end{document}